# Developing and Validating an Interactive Training Tool for Inferring 2D Cross-Sections of Complex 3D Structures

Anahita Sanandaji[1], Cindy Grimm[2], Ruth West[3], Christopher A. Sanchez[2]

[1]Ohio University, [2]Oregon State University, [3]University of North Texas

**ABSTRACT**

Understanding 2D cross-sections of 3D structures is a crucial skill in many disciplines, from geology to medical imaging. Cross-section inference in the context of 3D structures requires a complex set of spatial/visualization skills including mental rotation, spatial structure understanding, and viewpoint projection. Prior studies show that experts differ from novices in these, and other, skill dimensions. Building on a previously developed model that hierarchically characterizes the specific spatial sub-skills needed for this task, we have developed the first domain-agnostic, computer-based training tool for cross-section understanding of complex 3D structures. We demonstrate, in an evaluation with 60 participants, that this interactive tool is effective for increasing cross-section inference skills for a variety of structures, from simple primitive ones to more complex biological structures.

## 1. INTRODUCTION

The ability to understand 2D cross-sections of 3D structures plays an important role in many scientific domains including medical imaging, biology, geology, architecture, and engineering [11, 21, 24, 28, 29, 47]. Identifying and orienting the correct 2D cross-section of a 3D object requires a complex set of spatial/visualization processes, including mental rotation, spatial structure understanding, and viewpoint projection. For example, in medical imaging (in particular, 3D volume segmentation) the segmenter operates on 2D cross-sections of the 3D data, mentally integrating contours drawn on these slices into a coherent 3D anatomical structure. Results from the in-depth field studies on segmentation with human experts indicate that they have a better understanding of the expected 2D cross-sections of 3D structures — and tend to use 3D views more frequently — than novices [47, 62]. Based on this observation, we hypothesize that training spatial reasoning skills (independent of the actual segmentation interface) would especially benefit novice segmenters.

The ability to infer a cross section of an object is positively correlated with spatial visualization ability [11, 12, 29, 30]. Spatial abilities represent a discrete set of cognitive processes that are tasked with the understanding and internalization of visuospatial information from the real world [34]. These faculties have been shown to be especially important in reasoning with spatial information and completing complex tasks that have a large intrinsic spatial component [23]. Importantly, these spatial abilities do naturally vary across the population [26], and there has been some suggestion that this natural variance may underlie or influence achievement within highly spatial fields such as science and engineering [59]. Individuals with lower spatial ability might therefore be at a disadvantage in certain scientific fields, including 3D volume segmentation.

Fortunately, there is ample evidence in the literature that suggests spatial skills can be improved through experience and training [3, 9, 12, 13, 25, 32, 33, 48, 53, 60]. For example, studies show that playing video games like "Tetris" or first-person shooter (FPS) games can improve performance on spatial ability tests [10, 17, 40, 48, 57]. Despite this ample evidence in the literature, questions remain about how to effectively train 2D cross-section understanding of non-trivial 3D structures, and how we can use computing systems and interactive user interfaces to implement such a training.

In this paper we focus our attention on training the specific spatial skills necessary for inferring 2D cross-sections of complex 3D objects. Cross-section inference is challenging because it requires the application of multiple spatial processes, including encoding the spatial characteristics of the structure, imagining slicing the object, removing the section of the sliced object and the cutting plane, and creating an image of the cross-section of the object.

Previous research introduced successful training to improve 2D cross-sectional skills (based on drawing, passive examples, and correct/incorrect feedback) [12, 19]. However, these trainings are largely paper-based, use simple geometric shapes, and focus only on correctly matching a given object and plane to the corresponding 2D cross-section. For more comprehended tasks such as 3D image segmentation, users must also incorporate viewpoint selection to understand how the cross-section changes as the plane moves.

Also, previous studies show it is difficult to infer cross-sections of complex 3D structures without user control over the viewpoints and objects [28, 47, 62]. To our knowledge, there is no training tool--let alone a computer-based one-- to support user control and interaction in this context. Our novel approach takes advantage of interactive, computer-based, spatial training methods to provide more targeted, task-based training.

We designed our training tool based on a hierarchical model that captures difficulty levels for the different required spatial skills, such as viewpoint manipulation. The training tool is interactive and steadily increases the difficulty level of individual spatial skills (viewpoint, orientation of the slicing plane, etc.) through a series of guided tasks. We use both geometric and organic (e.g., structures with branching) 3D models in the training to more closely match the segmentation task. Within each guided task, participants infer the 2D cross-section of a 3D object while interacting with 3D models (viewpoint change), observing how a cross-section shape changes after adjusting the slicing plane location and/or orientation. The participants can request help as needed and receive corrective feedback before moving on to the next task.

The goal of our training tool is to help novices and people with lower spatial skills to independently understand and identify 2D cross sections of 3D structures using an interactive, task-based digital tool. Our main research questions are:

- **RQ1)** Does an interactive, tasked-based training tool effectively help people infer 2D cross-sections of complex 3D structures, independently?
- **RQ2)** What specific spatial visualization skills are enhanced after using our interactive training tool for inferring 2D cross-section of 3D structure?
- **RQ3)** Which training tool features are most effective in enhancing performance?

Our work is contributing to the literature by developing and evaluating a novel training application that is under user control and enhances skills for inferring cross-section of non-trivial structures in a short window of training. Adding visual interaction/animation does not always facilitate learning and can sometimes harm [8, 14, 22, 49, 50]. As our tool uniquely blends animation and interaction, its effectiveness should be carefully evaluated. The null control is necessary to make sure the observed improvements are not because of learning effects. Using controlled studies, we evaluated the effectiveness of our training tool by measuring spatial ability performance of the participants before and after the training.

**Contributions of this paper include:** 1) Developing a task based training tool and instantiating it in an interactive 3D user-interface. 2) Conducting a user study to show that the interactive training tool effectively helps people to infer 2D cross-sections of 3D structures. 3) Identifying the 3D spatial skills that are enhanced after using our interactive training tool.

## 2. TRAINING TOOL DESIGN AND MOTIVATION

Our interactive training tool consists of a sequence of tasks with increasing levels of spatial difficulty. Each task requires the participant to move and/or orient the cross-section plane in order to achieve a specific goal, which is described verbally (e.g., position the plane so it creates a cross section across the thinnest part of the object). Participants can interactively move the plane, change the viewpoint, show the current cross-section, request help, and receive feedback on the task.

We first discuss the observations and data analysis that motivated the structure of the tasks and the choice of stimuli. Next, we provide detail on the chosen tasks and what spatial skill each address. Finally, we discuss the UI design and the interactive elements (viewpoint, plane selection, cross-section, help) within the training tool.

**Background and Motivation**

Results from field studies [47, 62] suggest that expert segmenters have a better understanding of the expected 2D cross-sections of 3D structures. To conduct an accurate segmentation, it is not sufficient to just passively view what a 3D structure and its cross-section look like; expert segmenters also actively search for cross-sections that meet certain criteria [28]. They do this by spending a lot of time moving the virtual camera and the slicing plane around to better understand the current 3D structure. Previous works show that, compared to passive viewing, the active interaction with physical/virtual models leads to performance gains and transfer in the context of spatial visualization training [6, 15, 35, 41, 54, 56]. Specifically, working with virtual models is beneficial for learning the spatial configuration of complex 3D objects, such as organic structures [9, 20, 32] and may be particularly beneficial to individuals with lower spatial ability [27, 38].

In addition, as shown in [44–46], compared to static images, interaction (even just spinning the object) can improve users' ability to correctly visualize the 3D structures and cross-sections. However, we need to keep in mind that being able to view a complex 3D structure from any viewpoint may overwhelm novice (and even expert) users. Free viewpoint introduces so much complexity that it swamps the training task itself [28, 30–32]. For this reason, we incorporate viewpoint control, but in a very controlled fashion.

Animation that combines static images into a dynamic visualization and allows users to perceive continuous change is an option for spatial training [5, 36]. However, encoding transient dynamic images can be challenging for users. The size, location and color of objects, and direction of movement in an animation can direct users' attention away for the training task [36, 37]. On the other hand, interactive interfaces that permit users to pause, and restart a task on demand, can address some of the challenges posed by using animation [12, 51, 55]. It is also worthwhile to emphasize that complicated interfaces, with unnecessary features, can impair users' ability to encode key information in a training task [7].

[12] presents a short training intervention that uses interactive animation, integrated with drawing and feedback, to train participants to identify the 2D cross sections of 3D objects. While this training was effective for simple geometric objects, it does not support users working with more complex organic structures, searching for cross-sections, or moving the slicing plane or camera around to better understand the current 3D structure.

Considering all the above facts, we design and implement a training tool that: 1) Covers a broader range of tasks with different levels of difficulty that allows users to work with different organic 3D models. These tasks and models are similar to ones in a real segmentation process (e.g., working on structures with branching) 2) Provides interaction that allows users to move the virtual camera and the slicing plane around, see corresponding cross-sections on demand, and better understand the current 3D structure. 3) Controls free viewpoints to avoid overwhelming users. 4) Supports on demand "Help" options during the training and shows the correct answers at the end of each task. 5) Does not train users on exactly the same tasks/3D models they are tested on, therefore establishing whether or not the training transfers to new stimuli.

**Training Task Design**

We designed our training tasks to be goal-oriented and focused on the skills needed to understand the relationship between cross-section creating and viewpoint direction, slicing plane orientation, and 3D structure. Each task involves identifying where the desired cross-section is based on characteristics of the 3D structure, locally adjusting the plane to create a specific change in the 2D target cross-section or placing the plane to cut the 3D structure in a specific way. Designing the tasks was an iterative process and we developed many design alternatives.

To choose the tasks, we began by defining a set of canonical shape types (based on ones seen in the observational studies [47, 62]) and unique, identifiable cross-sections that arise from those shapes. We then chose four 3D objects to cover a range of object complexity: Two simple objects (an hourglass and a tapering shape), and two complex objects (branching Y shape and a potato shape with a hole). Figure 1 shows sketches from some of the initial set of shapes and tasks we considered. We created the final 3D models using 3D Studio Max [2], and ZBrush [43].

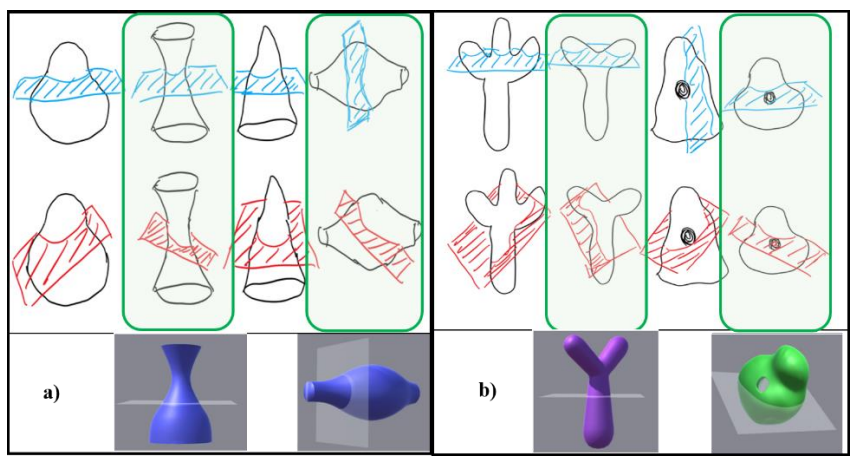

Figure 1: Two sets of canonical shapes with different slicing plane orientations. We selected two simple objects (an hourglass and tapering shape from set a) and two complex objects (Y branching and potato with hole from set b) to use in the training.

We characterized the difficulty for each spatial skill of each task, then culled the list, minimizing the number of times the shape changed and the total number of tasks while still covering three aspects of difficulty for each spatial task (object complexity, viewpoint orientation, and slicing plane direction). We organized these into three levels with 1-3 tasks per level. For each task we now describe the task and the corresponding visual skills it trains.

**Level 1:** Consists of three tasks with simple 3D objects (hourglass and tapering structures). The goal of level 1 tasks is to train participants to identify the cross-sections of 3D objects through simple plane movements/rotations. For all three tasks the participants must mentally cut the shape with the plane and rotate the resulting cut shape in order to visualize the cross-section.

- *Task 1 (difficulty=1): "Move the plane so that it creates a cross-section for the skinniest/ thinnest part of the given 3D shape".*

This task involves a simple plane movement. Plane orientation is orthogonal, and the 3D structure is a simple hourglass shape. The starting viewpoint with respect to the 3D structure is orthogonal. To accomplish this task, participants must move the plane up/down to the skinniest part. Changing the camera (up/down) is useful in order to verify the cross-section, however it is not necessary to rotate the camera to left/right.

- *Task 2 (difficulty=2): "Move the plane so that it creates a cross-section for the fattest/thickest part of the given 3D shape".*

This task increases the difficulty by placing the camera, so it is not orthogonal to the object and plane, requiring participants to perform a mental rotation of the scene-plus-plane (or move the camera). Plane movement is still a simple up/down movement. Changing the camera (up/down) is necessary in order to verify the correct cross-section, however it is not necessary to rotate the camera to left/right.
- *Task 3 (difficulty=3): "Adjust the plane to change the cross-section shape from a circle to an oval".*

This task involves a simple plane rotation. Plane orientation is initially orthogonal (horizontal), and the 3D structure is a simple hourglass shape. To accomplish this task, participants must perform a rotation of the plane *and* understand that doing so elongates the cross-section. Changing the camera (up/down, left/right) is useful in order to verify the correct cross-section.

**Level 2:** Consists of two tasks with a complex 3D object (branching Y shape). These tasks focus on training participants to understand the cross-sections of a more complex organic 3D object under the same conditions as the Level 1 tasks.
- *Task 1 (difficulty=4): "Adjust the plane so that it creates a cross-section that cuts across both branches but not the stem."*

This task is identical in setup to Level 1, Task 2, but participants must understand that a cross-section across the branches produces two circles. Therefore, changing the camera (up/down) is useful in order to verify the cross-section, however it is not necessary to rotate the camera to left/right.
- *Task 2 (difficulty=5): "Adjust the plane to create a single, oval-ish cross-section that crosses both one branch and the trunk."*

This task is identical in setup to Level 1, Task 3, but involves understanding both the elongation effect of rotation with the two cross sections. For this task, participants must also both rotate and move the plane. Changing the camera (up/down) is necessary in order to verify the correct cross-section. It is also useful to rotate the camera left/right to see the plane orientation with respect to the object.

**Level 3:** Consists of one task with a more complex 3D object (potato shape with hole). This tasks both increases the complexity of the 3D object and requires the participant to actively change the viewpoint.
- *Task (difficulty=6): "Adjust the plane so that the hole in the object creates a circular cross section (surrounded by an oval-ish cross section for the outside of the shape). Place the plane through the fattest part of the shape."*

This task also involves both plane movement and rotation. Plane orientation (with respect to the object) is initially orthogonal. However, the viewpoint with respect to both the plane and the 3D object is not orthogonal, therefore it is not possible to see the hole from the initial viewpoint. To accomplish this task, participants must change the viewpoint to see the hole, move the plane to the center of the potato shape, change its orientation to fully capture the hole, then mentally cut the shape and rotate it to visualize the cross-section. Therefore, it is necessary to change the camera (up/down, left/right) in order to verify the correct cross-section.

**Training Tool Application Design**

We designed the actual interface using a participatory design methodology [52] with five participants and in three phases: 1) Low-fidelity Paper prototype; 2) Higher-fidelity prototype in Balsamiq [4]; and 3) User interface implementation in Unity [58]. For each design phase we asked participants to perform one or more of the tasks and provide us with their feedback. We successfully elicited major suggestions for usability refinements, such as the appearance of the main window, visibility and consistency of buttons, more representative buttons to support affordance [39] (e.g., curved arrows for rotation), more visible feedback and help options, and removing interface elements that were not used in a task rather than asking participants to not to use them. Figure 2 shows these three phases. We now describe the UI elements of the training tool.

*Task UI*. Each task has two UI modes implemented (on separate pages): "Play" and "Solution". On the "Play" page, participants complete a given task (Figure 2 c). Upon completing a task, participants hit the "Complete Task" button and are redirected to the "Solution" page where they can hit the "Show Answer" button and see the step by step correct answer for the previous task (Figure 2 d). The controls are frozen, but they can hit the "Show Solution" button as many times as they wish. There is also a button that takes them to the next task. The motivation is that seeing the correct solution might be beneficial, but we did not want to force participants to see it. In the tutorial, participants were encouraged to use the "Show Answer" button for at least the first and last tasks.

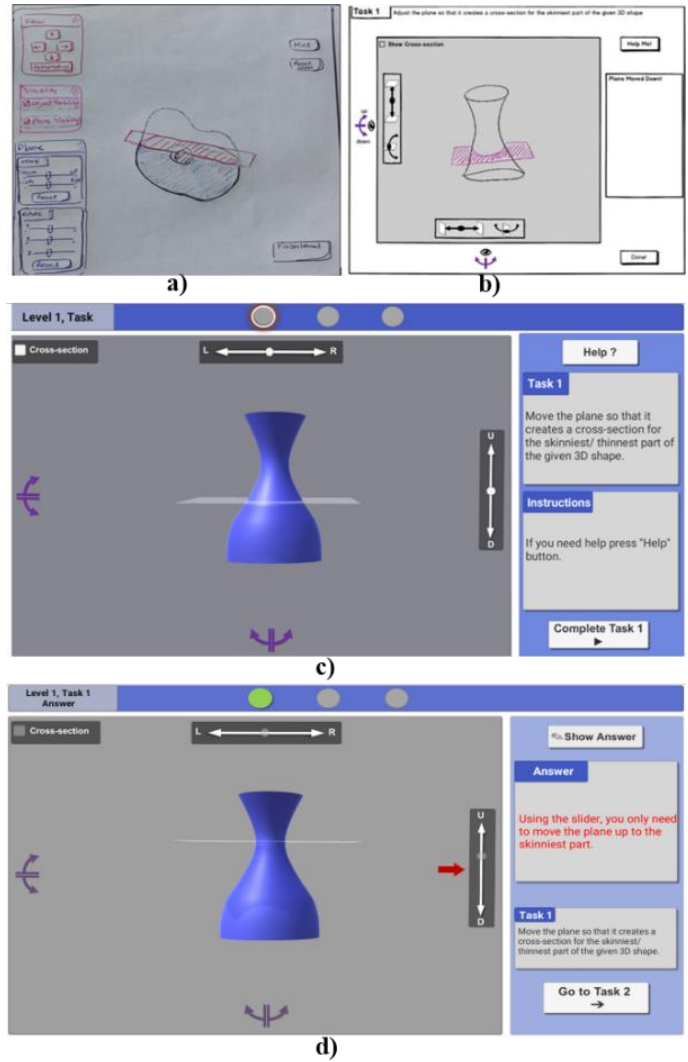

Figure 2: Sketches and prototypes: a) A low-fidelity sketch example; b) Balsamiq prototype; c) Unity Implementation: "Play" page, with plane rotation sliders hidden. d) Unity Implementation: "Solution" page for same task.

The UI for each mode has three separate panels: 1) Header panel: which has the level and task title along with a progress status image. 2) Main panel: which has the 3D model along with the buttons and sliders to move/rotate the plane, change viewpoint, and see cross-section changes. 3) Help and answer panel: which provides help options (in the "Play" mode) and shows answers (in the "Solution" mode).

*Viewpoint Visualization.* Viewpoint selection is an important spatial skill for understanding 2D cross-sections, both for finding object features and for determining the correct alignment of the plane with the feature. However, previous research shows that full viewpoint interactivity can confuse participants [28, 30, 31]. For our tasks, participants do not need full 3D viewpoint control; we only allow them to move the camera up and down or left and right using curved buttons. This still allows participants to view the object from all angles but keeps the object "up-right" in the view.

We characterize the viewpoint difficulty of our tasks by the relationship of both the object and the plane to the view direction. Orthogonal views of the object are easier to understand than non-orthogonal and placing the slicing plane perpendicular to the view direction means the cross-section is not distorted. We can increase viewpoint difficulty by placing the slicing plane so there is no way to view both orthogonally, and by starting with a viewpoint that is not orthogonal (forcing the participant to adjust the viewpoint) or where the desired feature is not visible (final task, see Figure 3).

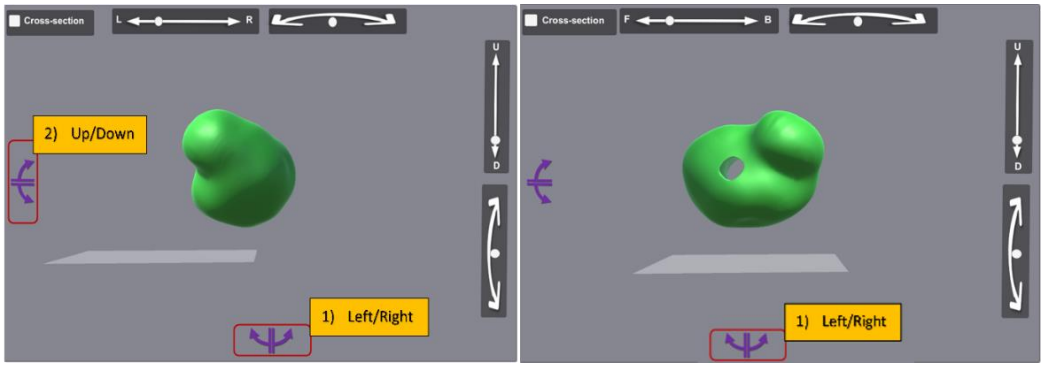

Figure 3: Level 3 Task. The hole is not visible in the initial viewpoint (left image). Using "Left" or "Right" buttons (labeled 1) we can rotate the object to change the view in order to see the hole (right image). Rotating the plane is required in this task so the plane rotation sliders are visible (unlike figure 2).

*Plane Movement and Rotation.* One of the skills needed for 2D cross-section understanding is to mentally adjust (rotate/move) the plane. To simulate this skill in our training tool, we added features to allow participants to move and rotate the plane. We included four sliders as UI elements for the plane movement/rotation. The representation and direction of "movement sliders" (straight) are different from the "rotation sliders" (curved) to increase affordances.

These sliders are only visible when they are needed to complete the task. For example, in Task 1 (Figure 2), participants only need to move the plane. Therefore, only the "movement" sliders are visible, and the "rotation" sliders are hidden. On the other hand, to complete Level 2 task 2, participants need to both rotate and move the plan, therefore all four sliders are visible (see Figure 4).

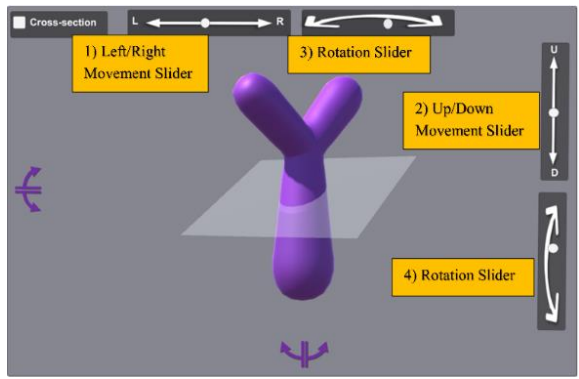

Figure 4: Level 2 Task 2 with all four sliders: "movement sliders" (straight arrows labeled 1 and 2) and "rotation sliders" (curved arrows labeled 3 and 4).

*Showing Cross-section from the Cut-away.* To identify a correct 2D cross-section, one should mentally cut the 3D structure and imagine the 2D representation from that cut-away. The difficulty of this task depends on the shape complexity of the 3D structure, the orientation of the slicing plane, and the viewpoint. In our tool, participants can see the cross-section by checking the "Cross-section" check box, which then shows the 3D structure "cut" open and the resulting cross-section rendered with a black and white texture. Figure 5 shows a top view of the cross-section for the potato shape with a hole (Task 3 Level 1).

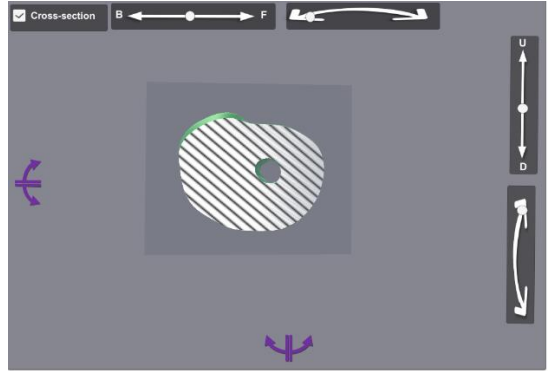

Figure 5: Level 3 task (potato shape, Figure 3), after selecting the check box and changing the view to see the model from the appropriate viewpoint. The correct cross-section is shown with a black and white texture.

*Help Button and Feedback.* To help participants accomplish the tasks we provided on-demand help options through a "Help?" button. This results in text instructions that provide more information on how to complete the task (such as "move the plane to the middle of the hourglass"). Help text is problem-dependent based on the location/orientation of the slicing plane and viewpoint. If participants create a correct cross-section and then click "Help?" button, the help text confirms that their answer is correct.

*Action Logging and Task Evaluation.* Every action that the participants take during the training session is logged by the tool application. We use the results of these logs for our analysis (e.g., how many times a participant requested help).

To evaluate participants' performance during the training, we logged elapsed time and score for each task. Successfully completing a task earns 100 points (total of 600). There was no penalty for wrong answers. The final state of each task is also saved as an image. The scores are not shown to the participants.

## 3. STUDY METHODOLOGY

In this section we describe the methodology used to evaluate our training tool and tasks. Our primary effectiveness evaluation was based on pre and post spatial evaluations that measured both 3D cross-section abilities and specific spatial skills (viewpoint, mental rotation, card rotation).

**Participants and Experiment Design**

We used a between-subject study design to evaluate our training tool. The treatment group used our training tool while the control group played the game Word Whomp [1] instead. In this game participants are asked to make words out of a group of six random letters to earn points. It requires approximately the same cognitive effort as spatial training but does not involve spatial skills. Word Whomp was previously used for the control group during a study evaluating the effectiveness of 3D games on enhancing 3D spatial abilities [48].

We recruited 60 participants (33 males, 25 females, and 2 who did not declare their gender) from the local community and university. The experiment session was conducted in groups of up to 5 participants at a time. We assigned each participant to the first available experiment session that met their time constraints and then randomly assigned each session to either the control or treatment condition. A total of 30 participants were randomly assigned to the control (Game) and 30 participants took part in the treatment (Training) group. Each study session consisted of three parts: pre-test, main training or game, and post-test.

*Spatial Ability Performance Measures and Tests.* We used the following four tests to measure spatial ability performance of our participants before and after the training. Three of the tests are standard spatial skill measures (3D rotation, viewpoint, and card rotation), while the fourth is a modified version of a standard cross-section skill test. The standard spatial skill measures were included in order to measure how much of the change in 3D cross-section skills could be attributed to changes in specific sub-skills.

*Mental Rotation Performance:* We measured 3D mental rotation performance by using an online version (implemented in Qualtrics) of the redrawn Vandenberg and Kuse [61] Mental Rotations Test (MRT) by [42]. The MRT consists of 24 questions and is divided into two sets of 12 items. Each question consists of five 3D block figures (see Figure 6 a). The target block figure on the top is compared to four similar blocks below. In each item, two of the four figures are rotated versions of the target (correct alternatives), whereas the other two are distractor figures. The task is to detect the two correct alternatives.

*Viewpoint Visualization Performance*: We used a modified version of Guay's Visualization of Views Test [16] to measure the viewpoint visualization ability. In this test, participants need to tell which viewing position a picture of a 3D object taken from (there are seven alternatives). This test is broken into two sets of 12 questions. Figure 6b shows a sample of the Guay's Visualization of Viewpoints test implemented in Qualtrics.

*Card Rotations (S-1):* This test is used to evaluate 2D spatial relation performance [18] and has two parts. Each part consists of a set of 10 items. Each item has nine 2D figures with the target shape at the left. Participants decide whether each of the eight other possible options represent either a rotation (choose "S" to confirm the option is the same as target) or a mirror image of the target (choose "D" to confirm the option is different from the target). We implemented this test in Qualtrics (see Figure 6c for an example item). We used part 1 in the pre-test and part 2 in the post-test.

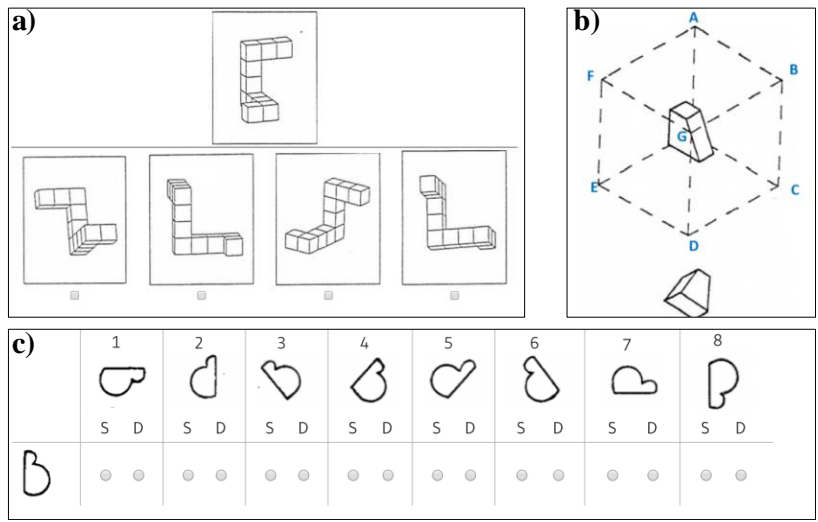

Figure 6: a) An example item from Mental Rotation Test (MRT) [61]. b) An example item from Guay's Visualization of Viewpoints test [16]. c) An example item from Card rotation S-1 test [18].

*2D Cross-section Understanding Performance [44]:* This test specifically measures cross-section understanding. There are three question categories:

- Category 1: Given a 3D structure and slicing plane, identify the correct 2D cross-section contour (Figure 7 a) and b).
- Category 2: Given a 2D cross-section for a 3D structure, identify the slicing plane that generated the contour (Figure 7c).
- Category 3: Given a 3D structure and multiple slicing planes, identify the valid contour sequence that corresponds to the slices (Figure 7d).

There are 12 questions of Category 1, six questions of Category 2, four questions of Category 3, and four controlling questions to check consistency. We used 13 questions for the pre-test and 13 questions for the post-test.

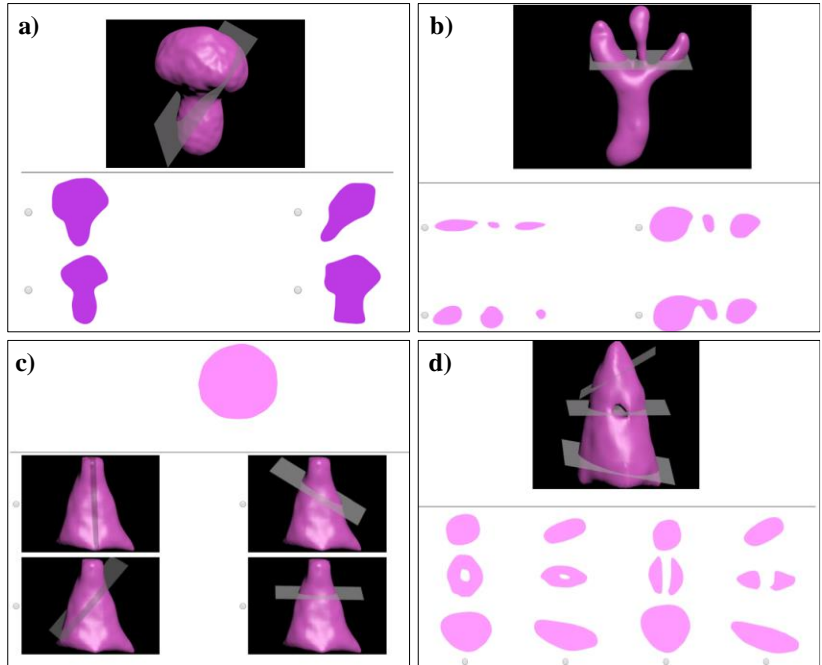

Figure 7: Test instrument for inferring 2D cross-sections [44]. Three examples of each question type: a) and b) Category 1. c) Category 2. d) Category 3.

*Demographic Questionnaire.* Our demographic questionnaire included: age, gender, level of education, level of experience working with 3D models and cross-sections, perceived benefit of the training, and a self-evaluation of success. We instructed participants to skip the questions rather than provide incorrect information if they did not wish to answer them.

**Procedure**

Each experiment consisted of three sub-sessions: pre-test, main training or game, and post-test. In the pre-test, all 30 participants completed the first part of our 2D cross-section understanding test. Our pilot studies showed that participants became extremely tired if they completed all the other three tests. Therefore, to avoid participants' fatigue, 10 of them were randomly assigned to complete one of the other tests: Mental Rotation Test (MRT); Visualization of View Test (VV); or S-1. Pre and post-test each lasted for around 20 minutes. The training group went through around the 30 minutes (time spent on training mean = 30.16 minutes with the SD = 2.13) of training while the control (game) group engaged in 30 minutes of playing the game.

During the training, a researcher introduced participants to the training tool application via a brief hands-on tutorial that explained how to use it. Participants then had around five minutes to explore the tool on their own. To avoid learning effects, the tutorial involved a task that was different from the main training tasks. The main training tasks followed the practice session. We asked our participants to complete all six tasks in the three levels as best as they could and spend some time on the "Solution" parts to understand how the cross-sections were created. Our application logged all participant interactions (e.g. button clicks) and computed their scores. We did not share the scores with our participants.

The post-test followed the main training/game session. We asked all of our participants to complete the second part of the 2D cross-section understanding test, and then the participants completed one of these corresponding tests (MRT, VV, and S1). Finally, they completed the demographic questionnaire. The entire study session took no longer than two hours.

## 4. RESULTS

We first compare the effectiveness results between the training and the game group. We next analyze the demographics of our participants and how that relates to pre and post-performance. We now discuss each of the findings here.

**Effects of Training Tool on Performance**

Table 1 shows descriptive statistics for all the test measures (pre and post), by test groups (training or game). As shown in the table, there were no significant differences in performance between the two test groups on any of the pre-test measures (all t-test values < 0.75, and all p-values > 0.31). Performance on the cross-section test was significantly correlated with MRT (rotation) ($r = 0.63$, $p = 0.0029$), and VV (viewpoint) ($r = 0.59$, $p = 0.0048$), but not S-1 (Card rotation) ($r = 0.43$, $p = 0.06$).

Table 1: Mean (SD) performance for each test measure by test group.

| Test Measure | Training | | Game | |
|---|---|---|---|---|
| | Pre | Post | Pre | Post |
| **2D Cross-section** | 9.1 (2.03) | 11.6 (1.53) | 9.6 (1.7) | 9.53 (2.4) |
| **MRT** | 17.3 (3.74) | 19.8 (3.34) | 17.7 (3.49) | 16.6 (4.7) |
| **VV** | 5.2 (3.09) | 7.5 (2.83) | 5.9 (2.96) | 6.2 (1.91) |
| **S1** | 71.8 (5.11) | 73.6 (3.23) | 72.55 (10.94) | 69.22 (14.61) |

Analyses of within-subject improvement from pre- to post test (paired t-test) for each of the test measures indicates:

- 2D cross-section test: Result for the training group is statistically significant ($t = 7.30$, $p < 0.0001$). For the game group the result is not statistically significant ($t = 0.15$, $p = 0.88$).
- Mental Rotation Test (MRT): Result for the training (pre and post) is statistically significant ($t = 3.55$, $p = 0.0062$). For the game group, the result is not statistically significant ($t = 0.56$, $p = 0.58$).
- Visualization of Views Test (VV): For the training (pre and post) results are statistically significant ($t = 4.44$, $p = 0.0016$). For the game group the results are not statistically significant ($t = 0.43$, $p = 0.67$).
- S-1 Card Rotation Test: Results for both the training and the game (pre and post) are not statistically significant ($t = 0.99$, $p = 0.3481$ and $t = 1.80$, $p = 0.11$).

In summary, test comparisons indicate that while there was no difference pre/post for the game group in all measures, there was a significant improvement for the training group in all test measures except for S-1 (card rotation). Figure 8 shows performance on all tests before and after the training/game.

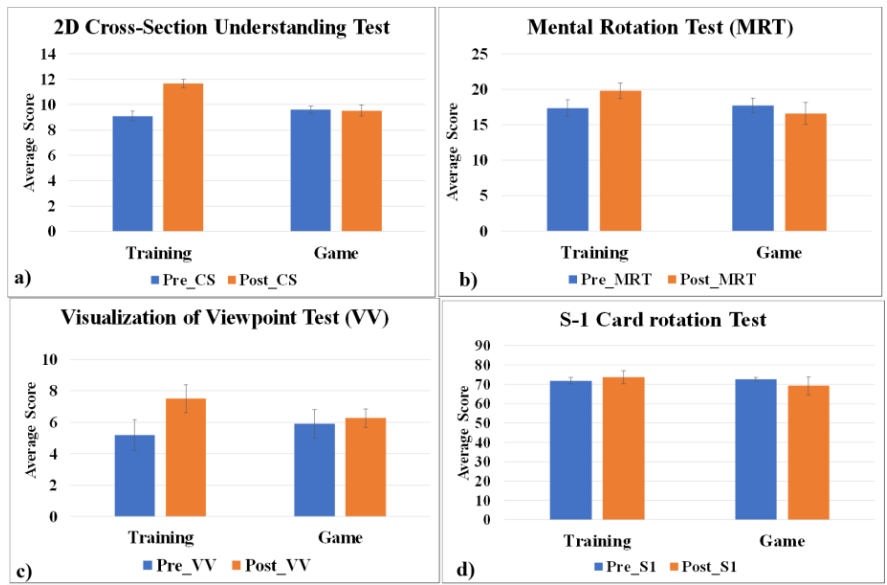

Figure 8: Difference in performance on each test measure for each group (training or game): a) 2D cross-section understanding test (maximum score for each part is 13); b) Mental Rotation Test (maximum score for each part is 24); c) Visualization of Views (maximum score for each part is 12); and d) Card rotation test S-1 (maximum score for each part is 80). Error bars represent the standard errors of the means.

There is also a significant negative correlation between the "2D Cross-section" pre-test performance and test performance improvements for the training group (r = -0.6974, p-value< 0.0001). Therefore, participants with lower 2D cross-section test scores show more improvements than the participants with higher scores. Figure 9 shows the average percentage of ratings for each of the demographic questionnaire for both the training and game groups.

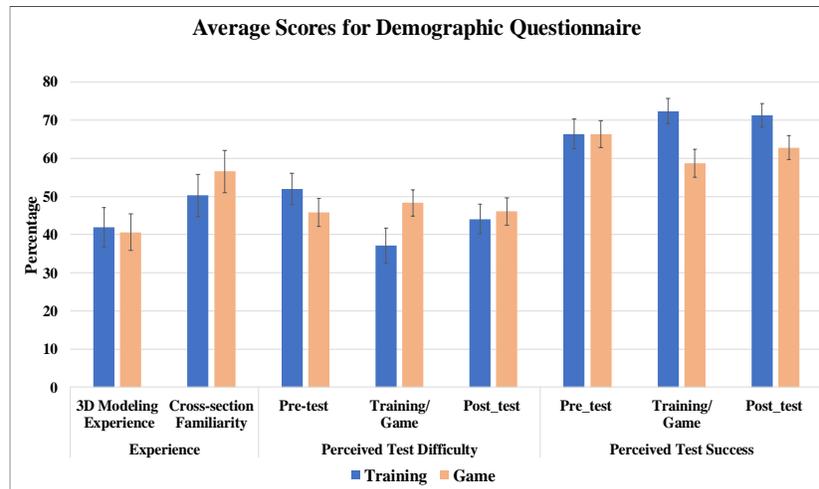

Figure 9: Answers demographic questionnaire. Training vs. Game group. Error bars represent the standard errors of the means.

Overall there is no significant difference between the training and game group in terms of experience. A simple one-way ANOVA shows there is a significant effect of 3D modeling experience on test performance for the pre-test: $F(1,57) = 14.47$, p = 0.00035. There is also a significant positive correlation between "2D cross-section pre-test" performance and "Cross-section Familiarity" (r = 0.3335, p = 0.0098). In addition, there is a significant negative correlation between "2D cross-section understanding test" improvements (difference of post-test and pre-test scores) and "Cross-section Familiarity" for the training group (r = -0.697, p<0.0001). The average perceived benefit of the test on enhancing the cross-section understanding (just for the training group) was 76%. There is a positive correlation between "2D cross-section test" improvements (difference of post-test and pretest scores) and "Perceived Benefit" of training (r = 0.2395) but the result is not significant.

**Training Tool Features and Test Performance**

During the training session, participants interacted with, and had access to, several features designed to improve their understanding: "ask for help", "move/rotate plane", "change view to up/down", "change view to left/right", "ask to show answer", and "check the cross-section". Table 2 shows the average training tasks scores along with the average frequency of interactions with each of the features of the training tool.

Table 2: Training tasks average score and average frequency of interactions with each of the features of the training tool.

|  | *Training Score* | Help | Move Plane | Rotate Plane | Change View Up/down | Show Answer | Check Cross-section |
|---|---|---|---|---|---|---|---|
| **Mean** | *449.33* | 15.66 | 746.4 | 740.06 | 35.9 | 4.76 | 24.63 |
| **SD** | *54.88* | 10.07 | 311.74 | 386.32 | 13.80 | 1.49 | 17.15 |

We analyzed the data for correlations between training tool features usage and training task scores. Table 3 summarizes these correlations. Except for the "Request to show answer", all of the other relationships are not significant. There is a significant negative correlation between the number of requests to see the task correct answers and pre-test scores. There is a positive correlation between requests to see the answer and test score improvements. This implies that participants who often requested to see the correct answer had lower scores for the 2D cross-section pre-test. Similarly, those participants with the most improvements also asked to see the correct answer more frequently.

There is also a significant negative correlation between average requests to show the answer and average score for each training task (r = -0.8621, p-value< 0.00001). This indicates that in general, participants mainly requested to see the correct answer for those tasks that they found more challenging, particularly Level 2 Task 2 (cross branch and tree-trunk) and Level 3 task (potato with hole).

Table 3: Training tasks average score and average frequency of interactions with each of the features of the training tool.

| Feature/score | 2D Cross-section Pre-test | Test Improvement |
| --- | --- | --- |
| # Requests for help | r=-0.1347, p>0.05 | r=-0.2245, p>0.05 |
| # Moving the plane | r=0.1525, p>0.05 | r=-0.2540, p>0.05 |
| # Rotating the Plane | r=0.1525, p>0.05 | r=-0.1667, p>0.05 |
| # Changing view to Left/Right | r=0.1525, p>0.05 | r=-0.2320, p>0.05 |
| # Changing view to Up/Down | r=0.1525, p>0.05 | r=-0.1228, p>0.05 |
| # Requests to show answer | **r= -0.4288, p=0.018** | **r= 0.4222, p=0.010** |
| # Check Cross-section | r=0.1525, p>0.05 | r=-0.2935, p>0.05 |

## 5. DISCUSSION

We now frame the results in terms of our original research questions.

*[RQ1, Training effective].* Results showed significant performance gains on inferring 2D cross-section for participants of the training group. We did not observe any significant pre/post-test score differences for the game group. Therefore, we can conclude that the tool does effectively improve participant's ability to infer 2D cross-sections of complex structures.

Participants with more 3D modeling/2D cross-section experience showed better performance on both pre and post-tests. In addition, there was a significant negative correlation between pre-test performance and test improvements, meaning that the training was significantly beneficial for those who obtained lower pre-test scores. That said, we still detect performance improvements for participants who got high scores in the pre-test. We can conclude that the test was beneficial for everyone, but particularly for individuals with lower spatial skills or less 3D modeling experience/cross-section familiarity.

*[RQ2, Spatial skills].* For the training group, we observed a significant performance improvement on the MRT and VV test measures, but not S-1. We also detected that performance on the 2D cross-section test was significantly correlated with the MRT and VV test. Since our training tool tasks involved multiple mental relations/transition and viewpoint changes, this result is not surprising. This contrasts with the S-1 tests, which are primarily in 2D and involve basic 2D transforms, which we do not employ in the training tool. Combined, these results indicate that our training tool is targeting the skills specific to 3D cross-sections, and not simply improving spatial skills across the board.

*[RQ3, Effective features].* Among all the training tool features, the functionality to "show correct answers" had the most significant impact on performance. Participants who more frequently requested to see the correct answer performed better in the post-test. This result confirms how crucial it is to provide participants with step-by-step correct explanations for specific tasks.

One of our predictions was to see a positive correlation between number of "help" requests and test performance improvements. However, we could not detect any significant relationship. One explanation is that the help feature was not that useful for our participants. One possible future direction is therefore to work on the "Help feature" so that it more closely matches the information gleaned from "show correct answers".

*Other Observations.* While the focus of this paper is not on the contribution of gender differences to 3D spatial ability performance, we conducted some relative analysis. Previous research results [11, 44] show that males significantly outperformed females in 2D cross-section understanding tasks. But in the current study, results indicate that there was no significant effect of gender difference on overall test performances (F(1,58) = 2.78, p = 0.1). Both females and males showed improvements after the training. So, the training was useful for both genders.

More than 80% of the participants were between 18-30 years old, but age does contribute on performance. Average score for the perceived benefit of the training tool was 76%. Also, on average, the training group participants gave lower scores for the perceived difficulty of the post-tests, and higher scores for the perceived success. According to all these results we claim that participants found the training to be useful.

In summary, while different methods of training are effective in enhancing 3D spatial abilities (e.g., [12, 48]), we argue that targeting the difficult skills and explicitly stair stepping learning provides better learning outcomes for understanding 2D cross-sections of more complex structures, given the same amount of invested time. Our interactive training tool allows users to actively view what a 3D structure and its cross-section looks like, easily search for cross-sections, and therefore better understand the current 3D structure. Both training tasks and some of the test items are similar to the ones in a real segmentation process. Specifically, instead of using geometric/simple structures, we used organic 3D models. We also created training tasks/3D models that were different from the one used in the test and could successfully demonstrate that performance was both gained and transferred to new stimuli after the training.

## 6. CONCLUSION

Understanding 3D structures through 2D cross-sections is a spatial task that appears both in 3D volume segmentation and many other scientific fields. Similar to other shape understanding and spatial tasks, performance on this task depends on different spatial skills which vary across the population and can be improved through training and practice.

We are — long term — interested in creating training tools for 3D volume segmentation. To put our 2D cross-section strategy into practice, in this paper we introduced our interactive training tool for 2D cross-section understating as a first step towards designing and implementing a domain agnostic training tool.

The results of the experiments demonstrated that our training tool was effective not only in enhancing 2D cross-section understanding, but also improving two other spatial skills: mental rotation and viewpoint visualization. In addition, it produced a verifiable effect across those skills, utilizing only a small window of training. While most research studies on this area implement hours of training, similar to [48] findings, our results also suggest that smaller training windows do have some utility, and future research could possibly explore the actual effectiveness of training durations.

*Future Work.*

While our study results prove that the training strategy was effective for engineering students (many of our participants were undergraduate and graduate students), it is hoped that it can also be adapted to the training in other contexts, such as with children or older adults. Likewise, it is of interest to explore these effects with more participants with diverse backgrounds (gender, age, field of study, level of education), to evaluate their performance, and subsequent effectiveness of the training, and in so doing get a better understanding of how this training might interact with other demographic differences.

The next step for this research is to use a customized version of the training tool in medical/research labs and evaluate its effectiveness in ongoing manual 3D volume segmentation tasks.